# Computer Assisted Parallel Program Generation

Shigeo Kawata
Graduate School of Engineering, Utsunomiya University, Japan.

## INTRODUCTION

Parallel computation is widely employed in scientific researches, engineering activities and product development. Parallel program writing itself is not always a simple task depending on problems solved. Large-scale scientific computing, huge data analyses and precise visualizations, for example, would require parallel computations, and the parallel computing needs the parallelization techniques. In this Chapter a parallel program generation support is discussed, and a computer-assisted parallel program generation system P-NCAS is introduced.

Computer assisted problem solving is one of key methods to promote innovations in science and engineering, and contributes to enrich our society and our life toward a programming-free environment in computing science. Problem solving environments (PSE) research activities had started to enhance the programming power in 1970's. The P-NCAS is one of the PSEs; The PSE concept provides an integrated human-friendly computational software and hardware system to solve a target class of problems. For example, a PSE generates a computer program automatically to solve differential equations (Boonmee et al., 1998a; Boonmee et al., 1998b; Fujio et al., 1998; Fujita et al., 2000; Gallopoulos, et al., 1991; Gallopoulos et al., 1994; Hirayama et al., 1988; Houstis et al., 1992; Kawata et al., 2000; Okochi et al., 1994; Rice et al., 1984; Umetani, 1985). In the PSE concept, human concentrates on target problems, and a part of problem solving, which can be solved mechanically, is performed by computers or machines or software.

The concept of the computer-assisted program generation has been opening the new style of the computer programing to reduce the programing hard task. Huge computer software may include errors and bugs. The errors or malfunction of the software infrastructure may induce uncertainty and accordingly serious accidents in our society (Einarsson, 2005; Kawata et al, 2012). The programing process tends to include mechanical parts, that means mechanically programmable parts. When the application area of the software is limited in a reasonable size, a part of the software would be mechanically generated. For example, scientific research-oriented programs would have a similar program structure depending on the numerical scheme. So the present PSE systems including P-NCAS provide a new powerful tool for the programing assistance toward a programing-free environment (Boonmee et al., 1998a; Boonmee et al., 1998b; Fujio et al., 1998; Fujita et al., 2000; Gallopoulos, et al., 1991; Gallopoulos et al., 1994; Hirayama et al., 1988; Houstis et al., 1992; Kawata et al., 2000; Okochi et al., 1994; Rice et al., 1984; Umetani, 1985).

P-NCAS supports scientists and engineers to generate parallel programs for problems described by partial differential equations (PDEs). P-NCAS presents a remarkable capability of visualization and steering of all the processes required for the generation of parallel programs. In P-NCAS users input problem description information including PDEs, initial and boundary conditions, discretization scheme, algorithm and also comments on the problem itself as well as the parallelization information. P-NCAS supports a domain decomposition method for the parallelization, and the SPMD (single program multi data) model is employed. In P-NCAS users can see and edit all the information through the visualization and editing windows. The program flow is also visualized by a Problem Analysis Diagram (PAD). Even through the flow visualization window users can modify the program flow. Finally P-NCAS outputs the corresponding parallel program in the C language, and at the same time a document for the program including the problem itself, the program flow, the PDEs, the initial and boundary conditions, the discretization method, the numerical algorithm employed and the variable definitions.

In this chapter first the PSE concept is briefly introduced, and then the details of the P-NCAS system is described for the parallel program generation support. Finally an example application result is presented including a load balancing result.

## BACKGROUND OF COMPUTER-ASSISTED PROBLEM SOLVING ENVIRONMENT (PSE) IN SCIENTIFIC COMPUTING

PSE is defined as follows: "A system that provides all the computational facilities necessary to solve a target class of problems. It uses the language of the target class and users need not have specialized knowledge of the underlying hardware or software" (Gallopoulos, Houstis & Rice, 1994). In computing sciences, we need computer power, excellent algorithms and programming power in order to solve scientific and engineering problems leading to scientific discoveries and development of innovative new products. The computer power and the computing algorithms have been developed extraordinarily, and have provided tremendous contributions to sciences, engineering and productions. On the other hand, the programming power has not been developed well, compared with the computer power and the algorithm power. The concept of PSE was initially proposed to support the programming power in science and engineering, and has been explored for decades.

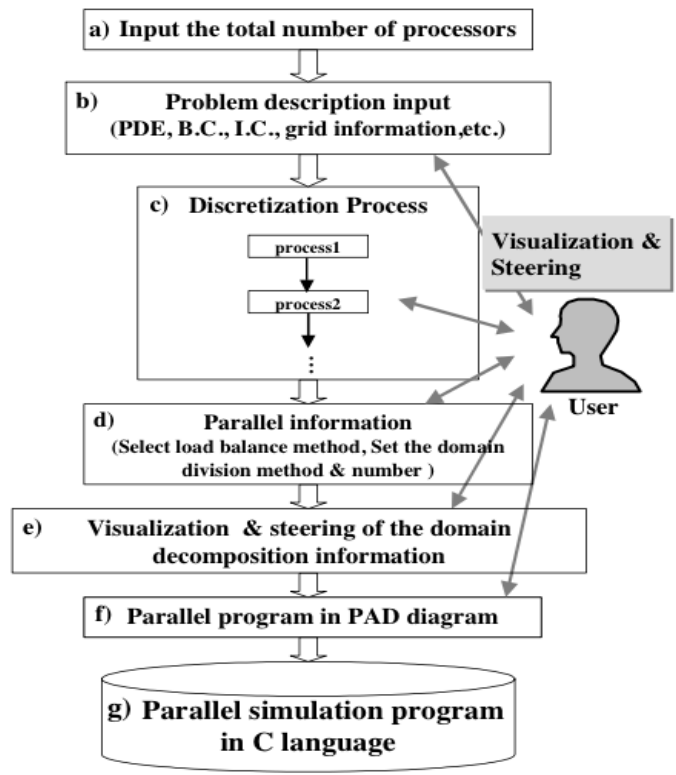

Figure 1. P-NCAS work flow for the computer assisted scientific parallel program generation support. P-NCAS inputs the domain-decomposition parallelization information, partial differential equations (PDEs), initial and boundary conditions, discretization method and algorithm, and outputs a C language program. The PDEs are automatically discretized and the program is generated mechanically. P-NCAS is a white box system, in which users can see and steer all the processes of program generation. P-NCAS system contains all the information for program generation, including basic equations, discretization schemes, discretized equations, boundary and initial conditions, parallelization scheme, mesh structure, program structure, and definitions of variables and constants. Therefore, a document for the corresponding program is also generated together with the program itself.

So far computer simulation has contributed to researches, productions and developments as well as experimental and theoretical methods. New researches tend to require new computer programs to simulate phenomena concerned. In developing new products engineers would also need new computer programs to develop new products effectively. They may have to develop the new programs or learn how to use the programs for the product development. Human power still contributes greatly to develop and write the new computer software. They like to devote themselves to solve their target problems, but not to develop or learn the computer programs. The PSEs would also help them develop the computer software including parallel programs or learn how to use the software system.

In these days PSE covers a various wide area, for example, computer-assisted program generation, education support, CAE (Computer Aided Engineering) software learning support, grid/cloud computing support, factory management support including plant factory management, program execution support, uncertainty management in scientific computing, etc. There are many PSE examples studied so far. In the references of (Ford & Chatelin, 1987; Fuju et al., 2006; Gaffney, & Houstis, 1992; houstis et al, 1997; Houstis, Rice, Gallopoulos & Bramley, 2000; Kawata, Tago, Umetani & Minami, 2005; Gaffney & Pool, 2007; Teramoto et al., 2007; Kawata et al., 2012) one can find the example PSEs.

## COMPUTER-ASSISTED PARALLEL PROGRAM GENERATION PSE - P-NCAS -

P-NCAS supports computer-assisted parallel program generation for PDEs-based problems in FDM (Finite Difference Method) based on a domain decomposition and MPI (Message-Passing Interface) functions (MPI, 2012). In addition, parallel programs generated in P-NCAS have load-balancing functions; one is a function for a static load balance and the other is that for the dynamic load balance. Figure 1 shows the P-NCAS work flow for the parallel program generation.

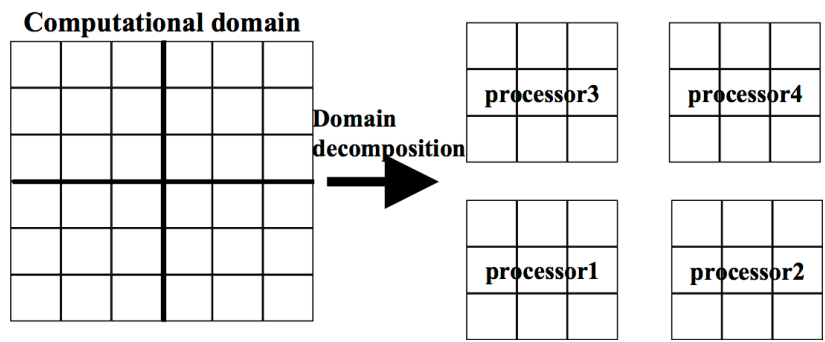

Figure 2. An example image of the domain decomposition in P-NCAS

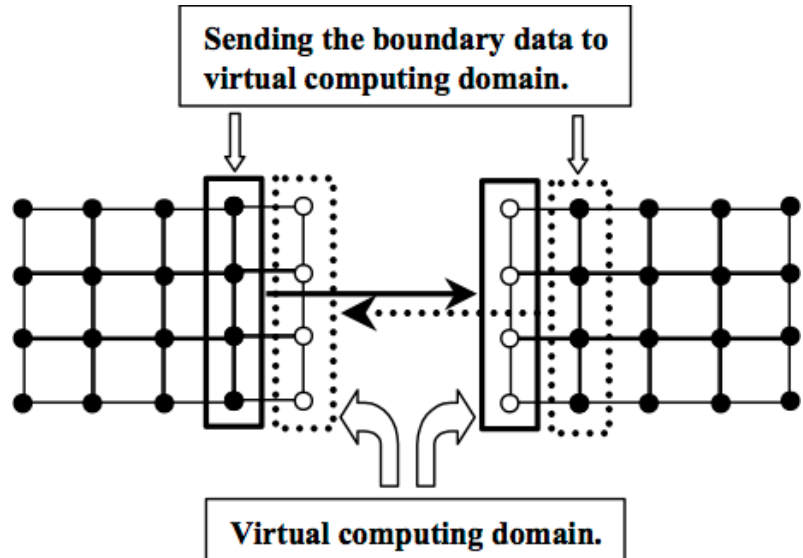

Figure 3. Boundary data are communicated by the MPI functions in P-NCAS, in which FDM is supported.

A PSE system of P-NCAS (Boonmee et al., 1998a; Boonmee et al., 1998b; Kawata et al., 2000; Fujita et al., 2000) inputs a problem information including PDEs, initial and boundary conditions, and discretization and computation schemes, and outputs a program flow, a C-language source code for the problem and also a document for the program and also for the problem. On a host computer a user inputs his/her problem, and P-NCAS guides the users to solve the problem. The P-NCAS contains all the information of the problem, PDEs, the discretization scheme, the mesh information, the equation manipulation method and results, the program structure designed, the variables employed, the constant definitions and the program itself. Therefore, P-NCAS also generates the corresponding document for the program generated together with the problem itself in P-NCAS (Inaba et al., 2004).

P-NCAS helps users generate MPI-based parallel simulation programs based on PDEs (Fujita et al., 2000). P-NCAS supports a domain decomposition in a design of a parallel numerical simulation program, and the domain decomposition is designed or steered by users through a visualization window. After designing the domain decomposition, the parallel program itself is designed and generated in P-NCAS, and the designed parallel program is visualized and steered by a Problem Analysis Diagram (PAD). In P-NCAS, MPI functions (MPI, 2012) are employed for the message passing, and a single program multiple data (SPMD) model is supported. The visualization and steering capabilities provide users a flexible design possibility of parallel programs.

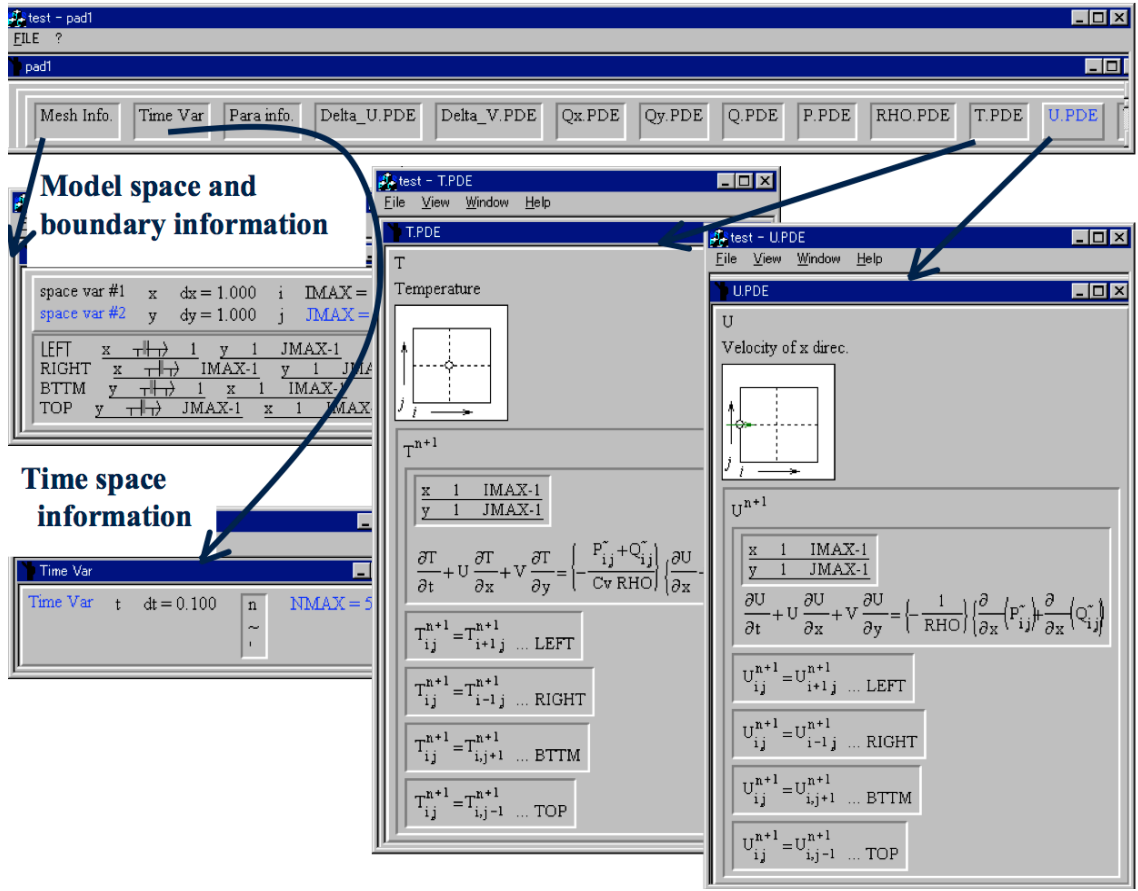

Figure 4. An example PDE-problem description in P-NCAS. Through each window users would edit the input description.

Figure 2 shows an example image of the domain decomposition in P-NCAS. For the data communication among the processors, the MPI functions are used. At least the boundary data for each domain decomposed are required to complete the computation in the adjacent processor (see Fig. 3). In P-NCAS the MPI functions are also automatically inserted to complete the parallel data communication programming. After specifying the domain decomposition information in P-NCAS, the parallel program is generated and provided to the users.

Figure 4 presents an example description of an input problem information, and Fig. 5 shows an example domain decomposition information. Through the P-NCAS visualization windows, for examples, shown in Figs. 4 and 5, one can check all the information and can also edit the information. In P-NCAS, after setting all the information for the problem description, the dicretization information and the parallelization information through the P-NCAS windows, all the information is visualized to the users and the users can edit all the information through the windows. The discretization of each PDE is also performed automatically; depending on the discretization information which users input through the P-NCAS windows, the PDEs are discretized and manipulated appropriately according to the PDEs solving scheme. Then P-NCAS designs the parallel program for the problem, and outputs the parallel program and the corresponding document. Figure 6 shows an example MPI program automatically generated in P-NCAS.

In addition, P-NCAS supplies users an option for a selection of a load balancing method, that is, an equal load balancing method or a dynamic load balancing method. When all the CPUs or machines are equivalent and the machine power is static, the equal load balancing method is appropriate. When the machines are heterogeneous or the machine power changes dynamically, the dynamic load balancing method is appropriate. Users can specify one of them, when they input the parallelization information through P-NCAS. In the dynamic load balance P-NCAS generates additional functions to measure all the machine power dynamically together with the corresponding computing program generated. The additional load balance measure functions measure the machine power regularly during the computation, and dynamically the domain sizes decomposed are modified to minimize the computation time.

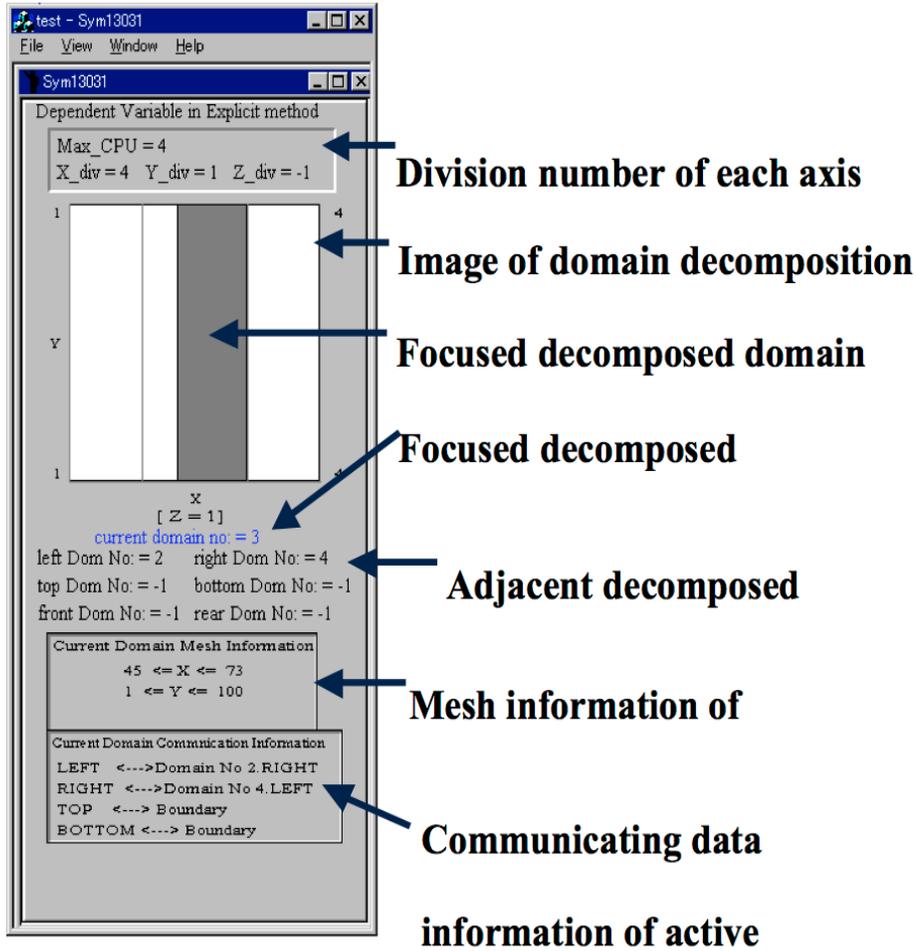

Figure 5. Input and visualization of domain decomposition information.

Figure 7 shows an example parallelization performance result for a shock wave propagation in a 2-dimensional compressible-fluid simulation: a parallel program generated in P-NCAS is used to measure the parallelization performance. In Fig. 7 the static load balance method is used on uniform machines. In order to check the dynamic load balance function automatically generated by P-NCAS, during the computation an additional load was applied as shown in Fig. 8 (the left graph): by the additional load the computation time increases much in this specific case, if the static load balancing is used. When the dynamic load balancing method is selected in this example case, P-NCAS generates the functions, which measure the load balance of each machine dynamically, and according to the measured result each domain size is changed and adjusted dynamically to minimize the computation time. The right graph in Fig. 8 demonstrates the viability of the dynamic load balancing functions generated in P-NCAS, and the computation time reduction is significant in this case.

# CONCLUSIONS

We have presented a concept of computer-assisted parallel program generation, and the P-NCAS system is introduced. The PSE concept and a brief history of PSE research activities are also presented in computing science. The PSE researches have explored capabilities of enhancement of programming power. PSEs have also started to provide a reliable tool to help engineers and scientists generate programs including parallel programs. There would not be an almighty PSE to solve all the problems, but each problem area would be solved by a specific PSE, for example, a PSE for a parallel program generation for PDEs-based problems by FDM and the domain decomposition; Another parallel program generation PSE would be needed for problems by FEM (Finite Element Method) or BEM (Boundary Element method) or particle methods, for example. A lot of PSE studies have demonstrated that each PSE in each area works well to support users to solve their target problems effectively. Even for uncertainty management in computing science PSE supports realistic tolls to avoid numerical errors or uncertainties (for example, Einarsson, 2005; Kawata et al., 2012).

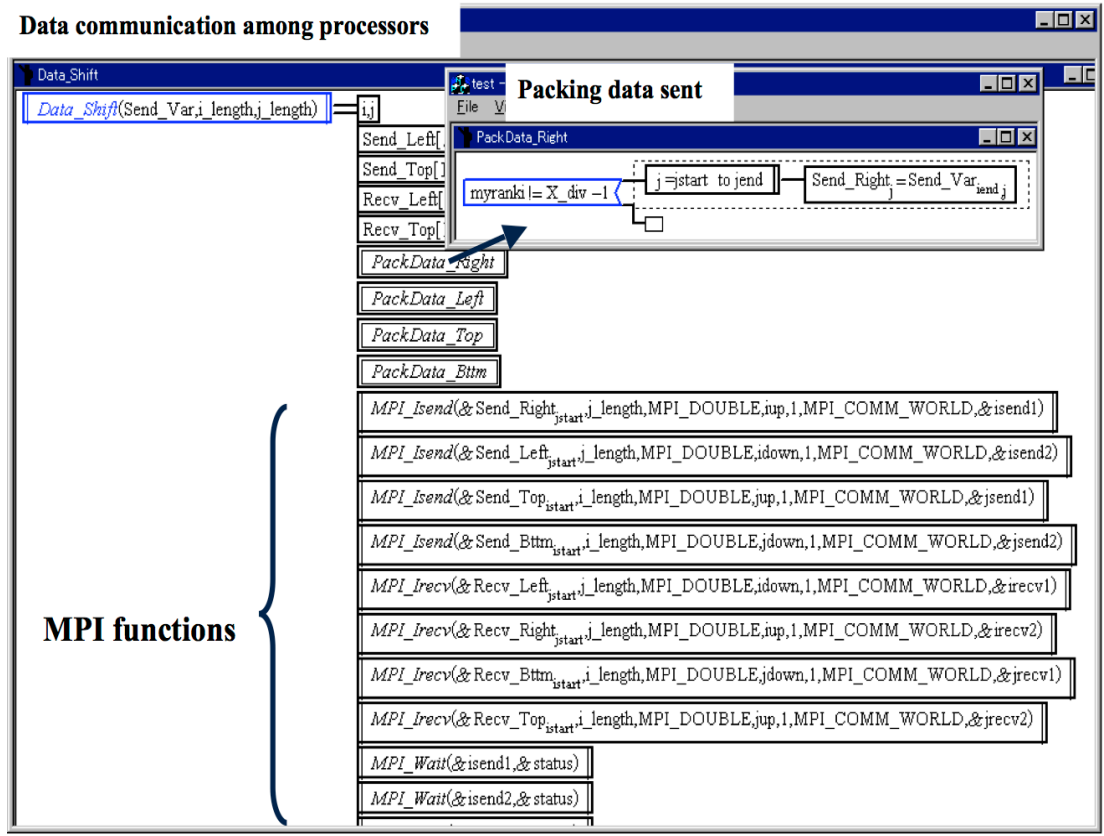

Figure 6. Visualization of MPI functions designed for a domain decomposition information in P-NCAS.

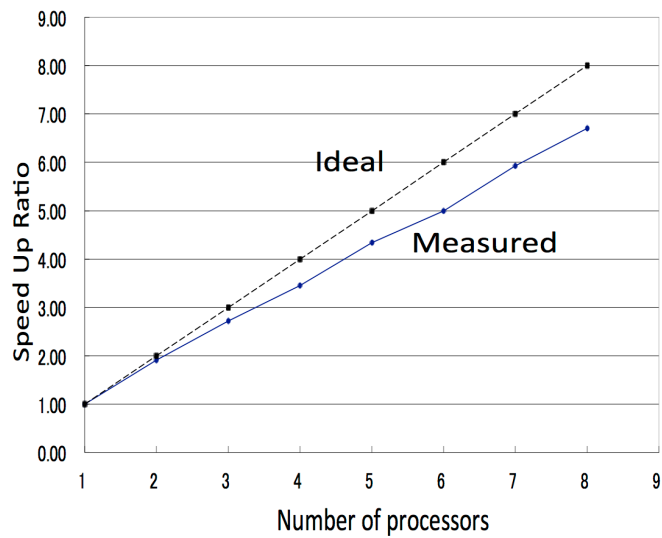

Figure 7. Speed up performance for a 2-dimensional fluid simulation (shock wave propagation) by a program generated by P-NCAS.

In this Chapter we presented P-NCAS for the parallel program generation support, and the results shown demonstrate that the P-NCAS, a parallel program generation support system, works very effectively to solve the problems in parallel. The results in P-NCAS show that the PSE concept and P-NCAS are viable for the problem solving in computing science. PSEs would be appropriate to enhance the programming power and to reduce the costs and heavy tasks in scientific computing as a reliable society infrastructure.

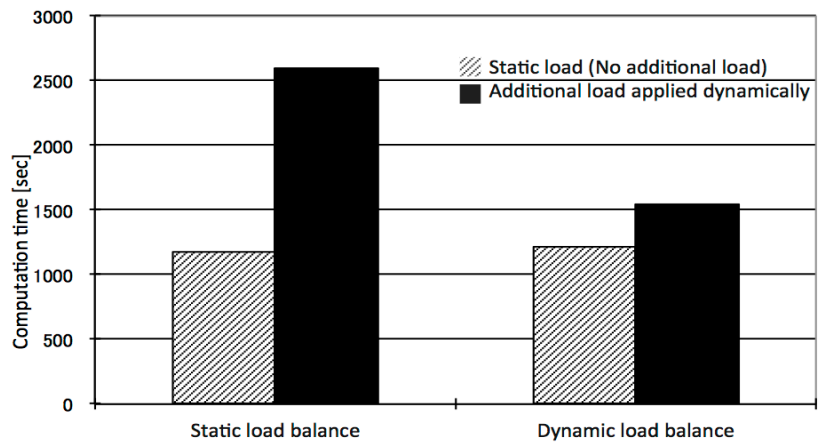

Figure 8. A performance test result for a dynamic load balance. An additional load was applied during the computation, so that the computation time increases as shown in the left graph. When the dynamic load balancing function generated automatically by P-NCAS is used to this specific case, the domain size changes automatically depending on the machine load, and the result of the dynamic load balancing in the right graph shows the viability of the dynamic load balance functions generated by P-NCAS.

## ACKNOWLEDGEMENTS

The work is partly supported by JSPS (Japan Society for the Promotion of Science), MEXT (Ministry of Education, Culture, Sports, Science & Technology in Japan), CORE (Center for Optical Research and Education, Utsunomiya University, Japan), Collaboration Center for Research and Development of Utsunomiya University, IFIP WG2.5, JSCES (Japan Society for Computational Engineering and Science) and PSE (Problem Solving Environment) Research group Japan. The authors also would like to express their appreciations to C. Boonmee, A. Fujita, T. Teramoto, H. Usami, Y. Manabe, Y. Hayase, M. Matsumoto, J. Rice, E. Houstis and friends in PSE Research group in Japan for their helps and fruitful discussions.

# REFERENCES

Boonmee, C., & Kawata, S. (1998a). Computer-Assisted Simulation Environment for Partial-Differential-Equation Problem, 1. Data Structure and Steering of Problem Solving Process. *Transaction of the Japan Society for Computational Engineering and Science*, Paper No. 19980001.

Boonmee, C., & Kawata, S. (1998b). Computer-Assisted Simulation Environment for Partial-Differential-Equation Problem: 2. Visualization and Steering of Problem Solving Process. *Transaction of the Japan Society for Computational Engineering and Science*, Paper No. 19980002.

Einarsson, B. (Ed.). (2005). *Accuracy and Reliability in Scientific Computing*. SIAM(Society for Industrial and Applied Mathematics).

Ford, B., & Chatelin, F. (Ed.). (1987). *Problem Solving Environments for Scientific Computing*. North-Holland.

Fujio, H., & Doi, S. (1998). Finite Element Description System as a Mid-Layer of PSE. *Proceedings of Conference on Computation Engineering and Science*, 3(2), (pp. 441-444).

Fujita, A., Teramoto, T., Nakamura, T., Boonmee, C., & Kawata, S. (2000). Computer-Assisted Parallel Program Generation System P-NCAS from Mathematical Model-Visualization and Steering of Parallel Program Generation Process-. *Transaction of the Japan Society for Computational Engineering and Science*, Paper No. 20000037.

Fuju, H., Kawata, S., Sugiura, H., Saitoh, Y., Usami, H., Yamada, M., Miyahara, Y., Kikuchi, T., Kanazawa, H., & Hayase, Y. (2006). Scientific Simulation Execution Support on a Closed Distributed Computer Environment. *the 2nd IEEE International Conference on e-Science and Grid Computing*, (27340112).

Gaffney, P. W., & Houstis, E. N. (Ed.). (1992). *Programming Environments for High-Level Scientific Problem Solving*. North-Holland.

Gaffney, P. W., & Pool, J. C. T. (Ed.). (2007). *Grid-Based Problem Solving Environments*. Springer.

Gallopoulos, E., Houstis, E., & Rice, J. R. (1991). Future Research Directions in Problem Solving Environments for Computational Science, Technical Report CSRD Report No.1259, *Report of a workshop on Research Direction in Integrating Numerical Analysis, Symbolic Computer, Computational Geometry, and Artificial Intelligence for Computational Science*, Washington DC.

Gallopoulos, E., Houstis, E., & Rice, J. R. (1994). Computer as Thinker/Doer: Problem-Solving Environments for Computational Science. *IEEE Computational Science and Engineering*, 1(2), 1-23.

Hirayama, Y., Ishida, J., Ota, T., Igai, M., Kubo, S., & Yamaga, S. (1988). Physical Simulation using Numerical Simulation Tool PSILAB. *The 1st Problem Solving Environment Workshop* (pp. 1-7).

Houstis, E. N., & Rice, J. R. (1992). Parallel ELLPACK, a development environment and problem solving environment for high performance computing machienes. In P. Gaffney, & E. N. Houstis (Ed.), *Programming Environments for High-Level Scientific Problem Solving*, North-Holland, (pp. 229-241).

Houstis, E. N., Gallopoulos, E., Bramley, E., & Rice, J. R. (1997). Problem-Solving Environment in Computational Science. *IEEE Computational Science & Engineering*, 4, 18-21.

Houstis, E. N., Rice, J. R., Gallopoulos, E., & Bramley, R. (Ed.). (2000). *Enabling Technologies for Computational Science, Framework, Middleware and Environments*. Kluwer Academic Publishers.

IFIP WG2.5 (International Federation for Information Processing, Working Group 2.5). (2012). IFIP WG2.5 homepage. Retrieved September 26, 2012, from http://www.ifip.org/wg-2.5

Inaba, M., Fuju, H., Kitamuki, R., Kawata, S., & Kikuchi, T. (2004). Computer-Assisted Documentation in a Problem Solving Environment (PSE) for Partial Differential Equation Based Problems. *Transactions of the Japan Society for Computational Engineering and Science*, 20040025.

Kawata, S., Boonmee, C., Fujita, A., Nakamura, T., Teramoto, T., Hayase, Y., Manabe, Y., Tago, Y., & Matsumoto, M. (2000) Visual Steering of the Simulation Process in a Scientific Numerical Simulation Environment –NCAS-. *Enabling Technologies for Computational Science*, E. N. Houstis, & J. Rice (Ed.), Kluwer, (pp. 291-300).

Kawata, S., Inaba, M., Fujiu, H., Sugiura, H., Saitoh, Y., & Kikuchi, T. (2005). Computer-Assisted Liaison among Modules in a Distributed Problem Solving Environment (PSE) for Partial Differential Equation Based Problems. *Transactions of the Japan Society for Computational Engineering and Science*, 20050029.

Kawata., S., Tago, Y., Umetani, Y., & Minami, K. (Ed.) (2005). *PSE Book: Computer assisted Problem Solving Environment in computing science* (Basic & Advanced) (in Japanese). Tokyo: Baifukan.

Kawata, S., Kobashi, H., Ishihara, T., Manabe, Y., Matsumoto, M., Barada, D., Hayase, Y., Teramoto, T., Usami, H. (2012). Scientific Simulation Support Meta-System: PSE Park - with Uncertainty Feature Information -. *International Journal of Intelligent Information Processing*, 3, 66-76.

MPI (Message passing Interface). (2012). *MPICH2*. Retrieved October 2, 2012, from http://www.mcs.anl.gov/research/projects/mpich2/

Okochi, T., Konno, C., & Igai, M. (1994). High Level Numerical Simulation Language DEQSOL for Parallel Computers. *Transaction of Information Processing Society of Japan*, 35(6), 977-985.

Rice, J. R., & Boisvert, R. F. (1984). *Solving Elliptic Problems Using ELLPACK*. Springer Series in Computational Mathematics 2, Springer-Verlag.

Teramoto, T., Okada, T., & Kawata, S. (2007). A Distributed Education-Support PSE System. *the 3rd IEEE International Conference on e-Science and Grid Computing*, (pp. 516-520).

Umetani, Y., (1985). DEQSOL A numerical Simulation Language for Vector/Parallel Processors. *Proceedings of IFIP TC2/WG22 (International Federation for Information Processing, Working Group 2.2)*, 5, (pp. 147-164).

## ADDITIONAL READINGS

The important references are listed in References above. The followings are especially recommended for the further readings on PSE (Problem Solving Environment): (Ford & Chatelin, 1987; Gaffney & Houstis, 1992; Gaffney & Pool, 2007; Houstis, et al., 2000) and on computer-assisted Parallel program generation: (Kawata, et al, 2012).

## KEY TERMS & DEFINITIONS

**Parallel Computing**: Computations performed in parallel on distributed computers or processors. Parallel computing program needs a special treatment to divide computations into many sub-computations on the distributed computers or processors. Special parallel computers would need special programing technique to divide the task into many sub-tasks or to achieve the communication among processors or computers if needed.

**Scientific simulation**: Computation to describe a real world in computer.

**PSE (Problem Solving Environment)**: Defined as "A system that provides all the computational facilities necessary to solve a target class of problems. It uses the language of the target class and users need not have specialized knowledge of the underlying hardware or software". An example typical PSE generates a computer program automatically to solve differential equations. PSE provides integrated human-friendly computational services and facilities. At present PSE covers a rather wide area, for example, program generation support PSE, education support PSE, CAE software learning support PSE, grid/cloud computing support PSE, job execution support PSE, e-Learning support PSE, uncertainty management in scientific computing, and PSE for PSE generation support. A parallel computing support PSE, P-NCAS is introduced in this Chapter.